\documentclass[10pt, twocolumn]{article}
\usepackage[text={6.5in,9in},centering]{geometry}               
\usepackage{mathptmx}  

\usepackage{graphicx}
\usepackage{amssymb}
\usepackage{epstopdf}
\usepackage{amsthm}
\usepackage{amsmath}
\usepackage{multirow}
\usepackage{tikz}
\usetikzlibrary{arrows}
\usepackage[strict]{changepage}

\usepackage{leading}
\leading{12 pt}

\usepackage{ifthen}
\usepackage{leading}
\leading{12pt}

\newcommand{\Var}{\mathrm{var}}

\newcommand{\expect}{\mathbb{E}}
\newcommand{\RAID}[1]{RAID\,{#1}}

\title{On the Reliability of RAID Systems: \\ An Argument for More Check Drives}
\author{Sarah Edge Mann, Michael Anderson, and Marek Rychlik}
\date{}

\begin{document}
\maketitle

%
%
%
%

\begin{abstract}
In this paper we address issues of reliability of RAID systems. We
focus on ``big data'' systems with a large number of drives and
advanced error correction schemes beyond \RAID{6}. Our RAID paradigm is
based on Reed-Solomon codes, and thus we assume that the RAID consists
of $N$ data drives and $M$ check drives. The RAID fails only if the
combined number of failed drives and sector errors exceeds $M$, a
property of Reed-Solomon codes.

We review a number of models considered in the literature and build
upon them to construct models usable for a large number of data and
check drives. We attempt to account for a significant number of
factors that affect RAID reliability, such as drive replacement or
lack thereof, mistakes during service such as replacing the wrong
drive, delayed repair, and the finite duration of RAID reconstruction. We
evaluate the impact of sector failures that do not result in drive
replacement.

The reader who
needs to consider large $M$ and $N$ will find applicable mathematical
techniques concisely summarized here, and should be able to apply them
to similar problems.  Most methods are based on the theory of
continuous time Markov chains, but we move beyond this framework when
we consider the fixed time to rebuild broken hard drives, which we
model using systems of delay and partial differential equations.

One universal statement is applicable across various models:
increasing the number of check drives in all cases increases the
reliability of the system, and is vastly superior to other approaches
of ensuring reliability such as mirroring.

\end{abstract}


\section{Introduction \label{sec: intro}}
RAID technology (see \cite{Patterson_1988}) has a single primary
focus: to apply mathematical techniques to organize data on multiple
data storage devices such that in the event of one or more device
failures, the original data stored is still available.  In recent
years, this feature has become much more important. The reason for
this is the rapid erosion of the relative reliability of data storage
devices.

For example, according to the manufacturer, a typical high capacity
disk drive will experience an unrecoverable read error in every 1 in
$10^{14}$ bits, as discussed in \cite{Panzer_2007}. The same drive can 
transfer data at a rate of $6
\times 10^{8}$ bits.  At this transfer rate, this data storage device
will lose data every $1.67 \times 10^{5}$ seconds, or roughly every 2
days.

To build high capacity storage systems, a hundred or more disk drives
may be deployed, combined in a single system. Using the manufacturer's
projections, this system would lose data every 30 minutes.  Hence, the
imperative for reliable mathematical techniques to recover from this
data loss are becoming increasingly urgent for systems that store
``big data.''

\ifthenelse{\boolean{false}}{
Another important but often overlooked characteristic of the
reliability of RAID systems is that they are usually constructed from
devices that were manufactured at the same time, in the same place,
using the same equipment and with components from the same suppliers.
They were shipped together, installed together, and share the same
physical cabinet, power and cooling. Hence, a single undetected defect
or mechanical, electrical or temperature over-stress in any of these
processes or components can lead to groups of failures occurring
together. The traditional reliability model of RAID systems that
assume completely independent failures does not reflect what is
actually deployed in the real world. Academic references cite measured
failure rates of disk storage at 100 times or more the rates published
by manufacturers. There simply is no reliable way to accurately
project the failure rate of a particular group of disk drives in
advance.
}{
}

There exists another important class of failure that is
critical to overcome in order to ensure that the original data is
still available: silent data corruption. This
occurs when a storage device delivers incorrect data, and reports it
as correct. These events are well known and have been repeatedly
measured and documented in the industry. Not a single storage device
manufacturer supplies a specification as to how often these events
will occur.  A well designed RAID system, to be truly resilient in the
face of all these failure scenarios, must be able to detect and
correct as well as recover from a wide class of reported and
unreported error scenarios.

In this paper, we attempt to quantify the increased reliability that
is achieved by constructing RAID systems with more robust error correcting codes (ECC). Standard RAID ECCs are often termed \RAID{6}, and are
resilient in the face of two drives failing. However, for large
systems, these codes are simply not robust enough, especially in the
light of ``expected'' failures that occur every few minutes.  With only
two drive protection, service events must be scheduled quickly, or the
system runs a serious risk of permanent data loss. Employing
additional check drives allows for fewer service events that can be
scheduled with more flexibility.

Well known Reed Solomon ECC codes can extend the reliability of RAID
systems to tolerate many more reported failures and succeed in
delivering correct data even in the face of silent data corruption.  
Since there is no accurate way to project the underlying
reliability or correctness of the individual data storage devices, we
propose that employing a more resilient mathematical technique is
imperative to the design of future RAID systems intended to store ``big
data.''

\subsection{Measures of RAID Reliability}
The most common metric used to measure the reliability of a RAID system is the Mean Time to Data Loss (MTTDL), which measures the average time it takes for a given RAID system to experience a failure in which data is irrecoverably lost.  However, MTTDL can be difficult to interpret and somewhat misleading, as discussed in \cite{Plank_NoMDL}, \cite{Paris_2008}.  


Here we will focus on the Probability of Data Loss within a specified deployment time $t$  (PDL$_t$).  This measure is more useful to a user of a RAID system than MTTDL in that it allows the user to think in terms of the acceptable risk of losing data during the expected lifetime of the data storage system, and is a more nuanced measure than MTTDL. However, to provide easy comparison with results from other authors, we will discuss the MTTDL for our models, as well as the PDL$_t$.  Some authors have chosen to focus on how much data a RAID system expects to lose in a given deployment time.  We argue that any data loss is unacceptable and thus focus solely on whether or not data is lost, not how much.  



\section{Model 1: No Repair \label{sec: no repair}}
Let us start by exploring a particularly simple model of the reliability of a RAID system.  Although this model is not nuanced, it will allow us to develop the relevant mathematical techniques in a case where analytical solutions are both tenable and concise.  In future sections, we will build off this model to create more complex and realistic models of RAID reliability.   Such models are examples of birth and death  chains, which have been well studied in the mathematical literature; see \cite{Goodman_2006} and \cite{Thompson_1981}, for example. 

Consider a RAID system consisting of $N$ data drives plus $M$ check drives.  There are $T = N + M$ drives in the system, with a storage rate $N/T$.   RAID storage systems are such that the system can tolerate and recover from up to $M$ drive failures; if there are $M+1$ or more failures, data is irretrievably lost.  See \cite{Plank_RS_Tutorial} and \cite{Plank_Correction} for the mathematics of such systems.  

In this first ``no repair'' model, we assume that drives are never repaired or replaced; if a drive fails, the system continues to operate without it.  So long as no more than $M$ drives fail, the system is functional and all data can be read and new data can be written.  
Maintenance can be expensive, so engineering a system that will never need to be touched by human hands within some fixed deployment time might be a good design solution.  Thus although this model is simple, it is also realistic.

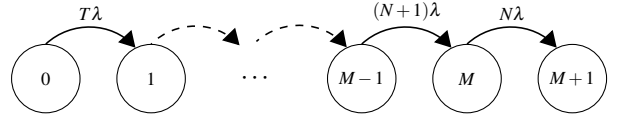
\begin{figure}[hbtp!]
\centering
\begin{tikzpicture}[scale =.7, transform shape, node distance = 2 cm, bend angle = 45, > = triangle 60]
	\tikzstyle{state}=[circle, draw, minimum size=13mm]
	\tikzstyle{fail}=[<-, bend right, semithick]
	\tikzstyle{recover}=[->, bend left, semithick]
	\node [state] 	(zero) 						{$0$} ;
	\node [state] 	(one)  		[right of = zero]		{$1$}
		edge 	[fail] 		node[auto, swap] {$T\lambda$} 	(zero.north);
	\node [circle, minimum size=14mm] 	(gap) [right of = one]{\Large{$\cdots$}}
		edge 	[fail, dashed] node[auto, swap] {} 			(one.north);
	\node [state] 	(minus one) 	[right of = gap] 		{$M-1$} 
		edge 	[fail, dashed] node[auto, swap] {} 			(gap.north);
	\node [state] 	(M) 			[right of = minus one] {$M$}
		edge 	[fail] node[auto, swap] {$(N+1)\lambda$}		 (minus one.north);
	\node [state] 	(plus one) 	[right of = M] 		{$M+1$} 
		edge 	[fail] node[auto, swap] {$N\lambda$} 	(M.north);	
\end{tikzpicture}
\caption{\label{fig: MarkovChainNoRepair} Markov chain for the no repair model of RAID reliability.}
\end{figure}
We model this system as a discrete state, continuous time Markov process with $M + 2$ states, as shown in Figure~\ref{fig: MarkovChainNoRepair}.   State $i$ indicates that $i$ drives have failed.  The system is initialized in state $0$ with all drives working.  When a drive fails the system moves from state $i$ to state $i+1$.  If drives fail independently at a constant rate of failure $\lambda$ per drive,  then the system moves from state $i$ to $i+1$ with an \emph{effective failure rate} $\lambda_i = (T-i)\lambda$. If the system enters state $M+1$, the \emph{failure state}, then the RAID system has failed, and data has been lost.  

The probability distribution
on the set of states is a probability vector
\[\mathbf{q}(t)=(q_0(t),q_1(t),\ldots,q_{M+1}(t))^T\]
where $q_j(t)$ is the probability that the system is in state $j$ at time $t$.
Thus, $\sum_{j=0}^{M+1}q_j(t)=1$.
The evolution of the probability distribution $\mathbf{q}(t)$ is governed
by the system of ordinary differential equations (the {\em
  Kolmogorov-Chapman equations}):
\begin{equation}
	\label{eq: Kol-Chap}
	 \frac{d}{dt}\mathbf{q} = A(t)\mathbf{q}, \qquad \mathbf{q}(0) = (1, 0 , \ldots, 0)^T 
\end{equation}
where $A(t)$ is the {\em transition matrix}.   For the no repair model, 
\begin{small}
\begin{equation}
A(t)=\left[\begin{array}{cccccc}
-\lambda_0 	& 0                      	& \ldots 	& 0        			&0		& 0\\
\lambda_0  	&-\lambda_1	& \ldots 	& 0        			& 0     	& 0\\
0          		& \lambda_1     & \ldots 	& 0        			& 0     	& 0\\
\vdots 		& \vdots 		& \ddots 	& \vdots 			& \vdots 	&\vdots \\
0  			& 0                      &\ldots 	& 0 				& 0 		& 0 \\
0  			& 0          		&\ldots  	&-\lambda_{M-1} 	&0 & 0 \\
0                          & 0                      & \ldots 	& \lambda_{M-1}       &-\lambda_{M} &0\\
0          		& 0                     & \ldots & 0                       		& \lambda_{M}            & 0
\end{array}\right]
\end{equation}
\end{small}
and $\lambda_i = -(T-i)\lambda$.   Notice that the eigenvalues of  $A$ are 0 and $-\lambda_i$ for $0 \leq i \leq M$.

If $X(t)$ is the state
of the system at time $t$ (so $X(t)\in\{0,1,2,\ldots,M+1\}$) then
the transition probability satisfies the equation:
\[P\big(X(t+h)=j \,|\, X(t)=k\big) = a_{jk}(t)\,h + o(h)\]
If the Markov process is stationary, the matrix $A(t)$ is constant; we will restrict our models to this case.
From the probability nature of the matrix it follows that
for all $t\ge 0$ $\sum_{j}a_{jk}(t)=0$.
The differential equation~(\ref{eq: Kol-Chap}) has solution
\begin{equation}
	\label{eq: gen state solution}
	\boxed{\mathbf{q}(t) = \exp(t\,A)\,\mathbf{q}(0)}
\end{equation}
where the matrix $\exp(t\,A)$ is the {\em matrix exponential}, i.e.
the sum of the series $\sum_{k=0}^\infty (t^k/k!)A^k$ (see \cite{Strang_1980}, p.~206).

\subsection{Calculation of PDL$_t$ \label{sec: pdlt calc}}

From the solution provided in (\ref{eq: gen state solution}), it is
easy to identify the PDL$_t$ for the model, it is simply the
probability of being in the failure state at time $t$: $\textrm{PDL}_t
= q_{M+1}(t)$.  For most models presented here, finding an analytical
form for this quantity will be intractable, or will result in an
expression that is too long to include.  However, the no
repair model is simple enough that we can tackle the PDL$_t$
calculation directly.


The calculations simplify significantly if we consider a slightly
modified model. The model consists of $T$ drives without partitioning
the system into the data and check drives. We run this system until
the last drive fails. However, formally this system is identical to
the original system with exactly $M'=T-1$ check drives and $N'=1$ data
drives. We consider the system failed when $M+1$ or more drives have
failed. Thus, in this new framing, the quantity of interest is:
\[
	\mathrm{PDL}_t= \sum_{i=M+1}^T q_i(t) =1-\sum_{i=0}^M q_i(t),
\]
and we find it by explicitly calculating $\mathbf{q} =
\exp(tA)\,\mathbf{q}(0)$.  The diagonalization of $A$,
$S^{-1}AS = D$, where $D$ is a diagonal matrix and $S$ is invertible,
can be found explicitly. Clearly, $D_{ii}=-\lambda_i=-(T-i)\lambda$ and
we found the following expression for the entries of $S$:
\[ 
	S_{kl} = (-1)^{k-l}{T-l \choose k-l}. 
\]  
Thus $S$ is a lower-triangular matrix whose entries are binomial
coefficients, up to the sign. The columns of $S$ are the right
eigenvectors of $A$.  The left eigenvectors of $A$ are the row vectors
which are the rows of $S^{-1}$. We found the following expression for
the entries of $S^{-1}$:
\[ S^{-1}_{kl} = {T-l\choose k-l} \]
This is again a lower-triangular matrix of binomial coefficients. 

When the matrix $A$ is diagonalizable, the explicit formula for $\exp(t\,A)$ is:
\begin{equation}
  \label{eq: Spectral-Decomp}
  \exp(t\,A) = \sum_{i=0}^T e^{t\lambda_i}\,\mathbf{v}_i\mathbf{w}_i^T
\end{equation}
where $\mathbf{v}_i$ is the $i$-th right eigenvector ($i$-th column of $S$) and $\mathbf{w}_i^T$
is the $i$-th left eigenvector ($i$-th row of $S^{-1}$). Since $\mathbf{q}(0) = \mathbf{e}_0=(1,0,0,\ldots,0)$, 
\begin{eqnarray*}
  q_k(t) &=& \sum_{i=0}^Te^{t\lambda_i}\left(\mathbf{e}_k^T\mathbf{v}_i\right) \left(\mathbf{w}_i^T\mathbf{e}_0\right)
  =\sum_{i=0}^Te^{t\lambda_i}S_{ki}\, S^{-1}_{i0} \\
  &=&\sum_{i=0}^T e^{t(T-i)\lambda}(-1)^{k-i}{T-i \choose k-i}\, {T\choose i}.
\end{eqnarray*}
We note the identity
\begin{eqnarray*}
{T-i \choose k-i} {T \choose i} &=& \frac{(T-i)!}{(T-k)!(k-i)!}\frac{T!}{(T-i)!i!} \\
&=&\frac{T!}{(T-k)!k!}\frac{k!}{(k-i)!i!}\\
&=& {T\choose k}{k\choose i}.
\end{eqnarray*}
This give the following:
\begin{eqnarray*}
  q_k(t) &=& {T\choose k} e^{-(T-k)\lambda t}
  \sum_{i=0}^T {k\choose i}\,(-1)^{k-i}\,e^{-(k-i)\lambda}\\
  &=& {T\choose k}e^{-(T-k)\lambda t}(1-e^{-\lambda t})^k.
\end{eqnarray*}
This is a binomial distribution $B(T,p)$, where $p=1-e^{-\lambda t}$ is the probability
of success, as expected, because we may consider the survival of each disk as an
independent Bernoulli trial. Hence,
\begin{equation}
  1-\mathrm{PDL}_t = \sum_{k=0}^M {T\choose k}e^{-(T-k)\lambda t}(1-e^{-\lambda t})^k.
\end{equation}
This formula is adequate for calculations and is numerically
stable. It may also be approximated by the left tail of the normal
distribution $N(T(1-p),\sqrt{T\,p\,(1-p)})$, based on the Central
Limit Theorem (CLT).  However, we can do better. Let $\mathcal{T}$ be
the time at which data is lost.  Then $F(t)=\mathrm{PDL}_t =
P(\mathcal{T} \le t)$ is the cumulative distribution function of
$\mathcal{T}$.  We find the p.d.f. of $\mathcal{T}$ using
$F'(t)=-\sum_{k=0}^Mq_k'(t)$.  Using the system of differential
equations satisfied by $q_k(t)$ ($q_0'(t)=-\lambda_0\,q_0(t)$ and
$q_k'(t)=-\lambda_k\,q_k(t)+\lambda_{k-1}\,q_{k-1}(t)$ for $k\ge 1$)
we find:
\begin{small}
\begin{eqnarray*}
  F'(t)&=&\lambda_0q_0(t)-\sum_{k=1}^M\left(-\lambda_k\,q_k(t)+\lambda_{k-1}\,q_{k-1}(t)\right)\\
  &=&\lambda_M\,q_M(t)=\lambda T\,{T-1\choose M}\,e^{-t(T-M)\lambda}\,(1-e^{-\lambda t})^M.
\end{eqnarray*}
\end{small}
Also, $F(0)=0$, and thus
\[ \mathrm{PDL}_t =\lambda T\,{T-1\choose M}\int_0^t e^{-t(T-M)\lambda}\,(1-e^{-\lambda t})^M\,dt.\]
We use the substitution $u=1-e^{-\lambda t}$, $du=\lambda (1-u)\,dt$, and obtain
\begin{eqnarray*}
  \mathrm{PDL}_t &=& T\,{T-1\choose M}\,\int_0^{u(t)} (1-u)^{T-M}\,u^M\,(1-u)^{-1}\,du\\
  &=&\frac{T!}{M!(T-M-1)!}\,\int_0^{u(t)} (1-u)^{T-M-1}\,u^{M}\,du.
\end{eqnarray*}
This means that the random variable $\mathcal{U}$ given by the transformation
$\mathcal{U}=1-\exp\left(-\lambda \mathcal{T}\right)$
has beta distribution
\begin{equation}
\label{eq: Beta-Pdf}
f(u; \alpha,\beta) = \frac{1}{B(\alpha,\beta)}
u^{\alpha-1}(1-u)^{\beta-1}
\end{equation}
with parameters $\alpha=M+1$ and $\beta=T-M$. We note that the
transformation is the c.d.f. of an exponential model. If $\mathcal{T}$
were exponentially distributed with this c.d.f. the resulting
$\mathcal{U}$ would have uniform distribution on the interval
$[0,1]$. This highlights the stark difference between the uniform time to
failure for a single drive, and the the time to data loss for a RAID
system, which has small spread around its MTTDL.

\subsection{Calculation  of MTTDL \label{sec: MTTDL calc}}
In this paper we will develop several methods for calculating MTTDL; the following method is based on the results of Section~\ref{sec: pdlt calc}.  The inverse formula
$\mathcal{T} = -\frac{1}{\lambda}\log(1-\mathcal{U})$
allows one to compute MTTDL. The variable $\mathcal{V}=1-\mathcal{U}$ is
beta distributed, with the parameters $\alpha$ and $\beta$ swapped. We find
from standard sources on beta distribution that
\[\expect\log(\mathcal{V}) = \psi(\alpha)-\psi(\alpha+\beta)=\psi(T-M)-\psi(T+1)\]
where $\psi$ is the \emph{digamma function}. Hence,
\[ \mathrm{MTTDL} = \expect\mathcal{T}
  =-\frac{1}{\lambda}\expect\log\mathcal{V}
  =\frac{\psi(T+1)-\psi(T-M)}{\lambda}.\]
The digamma function is the only solution of the functional equation $\psi(x+1)=\psi(x)+1/x$
monotone on $(0,\infty)$ and satisfying $\psi(1)=-\gamma$, where $\gamma$ is the Euler-Mascheroni constant.
Therefore, we can show easily that
\[\mathrm{MTTDL}=\frac{1}{\lambda}\sum_{k=T-M}^{T}\frac{1}{k}.\] 

Yet another method uses the widely known fact that $MTTDL=\int_0^\infty R(t)\,dt$,
where $R(t)=1-F(t)$ is the \emph{reliability function}, and it uses formula
(\ref{eq: Spectral-Decomp}). This calculation is left to the reader. A
more general, related approach, based on the Laplace transform is
presented in the next section.

\subsection{An Analytical Approach to MTTDL \label{sec: Analytic MTTDL}}
The {\em moments} of $\mathbf{q}(t)$ are the quantities
$m_k(\mathbf{q}) = \int_0^\infty t^k \mathbf{q}(t)\,dt$. These are
vector quantities, but typically we will only be interested in the last
component, $m_k(q_{M+1})$, as it is related to the distribution of the time of
failure.  In particular, MTTDL$\, = \expect \mathcal{T} = m_1(q_{M+1})$.
 Our exact analytical technique to calculate the moments is
based on the Laplace transform: $\widehat{\mathbf{q}}(z) = \int_0^\infty e^{-z\,t} \mathbf{q}(t) \,dt$.  We recall that the Laplace transform is the \emph{moment generating
function}: 
\begin{equation}
	\label{eq: laplace moment generating}
	\widehat{\mathbf{q}}(z) = \sum_{k=0}^\infty (-1)^k\frac{z^k}{k!}m_k(\mathbf{q}).
\end{equation}

We can obtain a second expression for $\widehat{\mathbf{q}}(z)$ by considering the resolvent of the matrix $A$: $R(z;A)=(z\,I-A)^{-1}$.
The quantity $z$ is complex, and $R(z;A)$ is a complex matrix-valued
function meromorphic in the entire complex plane. Its poles are
exactly the eigenvalues of $A$.
A well-known identity in operator theory relates the Laplace transform
to the resolvent and the initial condition:
\[\widehat{\mathbf{q}}(z) = R(z;A)\,\mathbf{q}(0).\]
The resolvent is a rational function, and in applications
to continuous-time Markov chains, $0$ is an eigenvalue of $A$.
The resolvent therefore has a pole at $z=0$ of some order $\nu$.
For this reason it is convenient to consider the Laurent series of
$R(z;A)$ at $z=0$:
\[ R(z;A) = \sum_{k=-\nu}^\infty z^k\,R^{(k)}.\]
The matrices $R^{(k)}$ are constant. When $z=0$ is a simple eigenvalue
(common case) then $\nu=1$ and $R^{(-1)}$ (the residue) admits an
expression in terms of the left and right eigenvectors of $A$ with eigenvalue 0.  Let
$\mathbf{u}^T\,A=0$ and $A\,\mathbf{v}=0$, scaled
so that $\mathbf{u}^T\,\mathbf{v}=1$. Then 
$ R^{(-1)}=\mathbf{v}\,\mathbf{u}^T $ is a \emph{spectral projection} on
the eigenspace of $0$. When $\mathbf{u}^T\,\mathbf{q}(0)=0$, as is the case in the models presented here, we have
\[\widehat{\mathbf{q}}(z) = \sum_{k=0}^\infty z^k\,R^{(k)}\,\mathbf{q}(0).\]

Comparing this equation with (\ref{eq: laplace moment generating}), we obtain an expression for the moments of $\textbf{q}$ in terms of the Laurent
series of the resolvent:
$m_k(\mathbf{q}) = (-1)^k\,k!\,R^{(k)}\,\mathbf{q}(0)$.
This is our main device to calculate means and variances of the time
of failure. These formulas are convenient to use with computer algebra
systems (CAS). 

The relationship between the moments of the time of failure $\mathcal{T}$ and
the resolvent can be explained easily. If $r$ is the initial state ($q_{j}(0)=\delta_{jr}$),
and the failure state is $s$, then the probability $F(t) = $PDL$_t$ of transitioning to the failure state before time $t$ is the entry $(e^{t\,A})_{s,r}$ of the
fundamental matrix. Therefore, the entry $R_{s,r}(z;A)=(R(z;A))_{s,r}$ of the resolvent is
the Laplace transform of the PDL$_t$. The function $F(t)$ is the c.d.f. of the time to failure.
The $k$-th moment of this distribution is $\int_0^\infty t^k\,dF(t)$. Stieltjes
integration by parts formula yields: 
$  m_k(\mathcal{T})=\int_0^\infty t^k\,dF(t) 
  = - \int_0^\infty t^k\,d(1-F(t))
  =-t^k(1-F(t))|_0^\infty+\int_0^\infty(1-F(t))\,k\,t^{k-1}\,dt
  = k\, m_{k-1}(1-q_{s,r}(t))$.
The Laplace transform of $1-q_{s,r}(t)$ is
$1/z-\hat{q}_{s,r}(z)=1/z-R_{s,r}(z;A)$.
Therefore, there is an explicit expression of the moments of the
time to failure in terms of the Laurent series of the resolvent:
\begin{equation}
  \label{eq: moment-fmla}
  \boxed{m_k(\mathcal{T}) = (-1)^k\,k!\,R_{s,r}^{(k-1)}.}
\end{equation}


\subsection{Summary of Results \label{sec: no repair results}}


In Section~\ref{sec: pdlt calc} we showed that if $\mathcal{T}$ is the random variable representing the time to failure, then the random 
variable \[\mathcal{U}=1-\exp(-\lambda\,\mathcal{T})\] is beta distributed
with parameters $\alpha=M+1$ and $\beta = N$. This provides a complete probabilistic
description of $\mathcal{T}$. We derived the explicit formula
\[\mathrm{PDL}_t =1- T{T-1\choose M}\, \sum_{i=0}^M\frac{(-1)^{M-i}}{T-i}{M\choose i}e^{-t(T-i)\,\lambda}.\]

In Section~\ref{sec: MTTDL calc} we derived an explicit formula for MTTDL:
\[\mathrm{MTTDL}=\sum_{i=0}^{M}\frac{1}{\lambda_i}=\frac{1}{\lambda}\sum_{k=T-M}^{T}\frac{1}{k}.\] 
We note that for large $T$ we have the following approximation:
\begin{eqnarray*}
  \mathrm{MTTDL} &=& \frac{1}{\lambda}\sum_{k=0}^M \frac{1}{1-k/T}\,\frac{1}{T}
  \approx\frac{1}{\lambda}\int_0^{M/T} \frac{1}{1-u}\,du\\
  &=&-\frac{1}{\lambda}\,\log\left(1-\frac{M}{T}\right)=\frac{1}{\lambda}\log\left(1+\frac{M}{N}\right).
\end{eqnarray*}
This formula demonstrates that we may increase
MTTDL to an arbitrarily large value by increasing the ratio
of check drives to data drives to infinity. 
The total number of drives, $T$, grows
exponentially with the target MTTDL. However, MTTDL grows linearly
with $M$ if $M\ll T$.

Thus, when there are relatively few check drives as compared to the
number of data drives, the economical way to double MTTDL is clear:
{\em double the number of check drives}.

\begin{figure}[hbtp]
	\centering
	\includegraphics[width = .45 \textwidth, viewport = 40 190 550 610, clip]{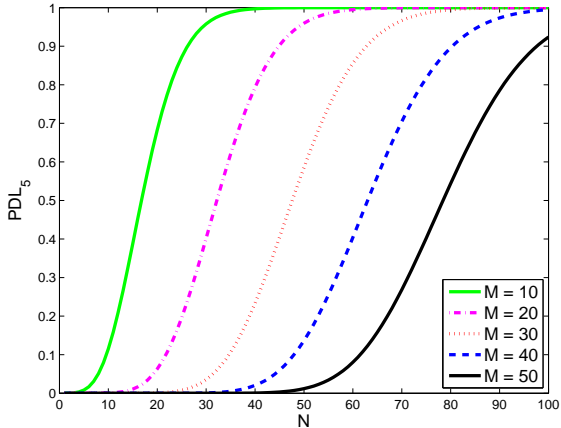} 
	\caption{\label{fig: no repair}PDL$_5$ for the no-repair model.}
\end{figure}

We can also numerically compute PDL$_t$ for this model using (\ref{eq: gen state solution}).    We will use the value $ \lambda = \frac{1}{10 \textrm{ years}}$ for numerical computation,
and assume that the expected deployment time for the system is five years. We believe this value for $\lambda$ is realistic, but perhaps conservative, based on the discussion of real world drive failure rates in \cite{Schroeder_2006}. We do not expect to obtain numerical results that are particularly exact, and thus will not dwell upon the value of such parameters used here.    Instead, we hope to better understand the effect of using additional check drives on the reliability of the system by examining numerical results qualitatively.

Figure~\ref{fig: no repair} shows PDL$_5$ under a no repair model as a function of $N$ for five values of $M$.  Notice that to maintain a particular level of reliability (PDL$_5$ value), more check drives are required as the number of data drives increase.


\section{Model 2: Individual Drive Repair} \label{sec: simple model}

Let us consider a model of RAID reliability as in Section~\ref{sec: no repair}, but now we allow failed hard drives to be repaired one at a time.  Other authors have also considered this model; see \cite{Plank_NoMDL}, \cite{Paris_2008}, \cite{Elerath_2007}, for example, or \cite{Burkhard_93} for a similar model.

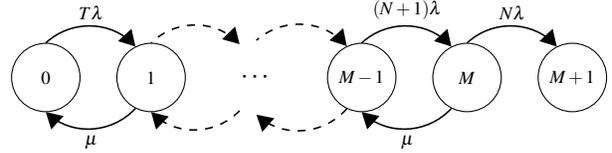
\begin{figure}[hbtp!]
\centering
\begin{tikzpicture}[scale =.7, transform shape, node distance = 2 cm, bend angle = 45, > = triangle 60]
        \tikzstyle{state}=[circle, draw, minimum size=13mm]
        \tikzstyle{fail}=[<-, bend right, semithick]
        \tikzstyle{recover}=[->, bend left, semithick]
        \node [state]   (zero)                                          {$0$} ;
        \node [state]   (one)           [right of = zero]               {$1$}
                edge    [fail]          node[auto, swap] {$T\lambda$}   (zero.north)
                edge    [recover]       node[auto] {$\mu$}                              (zero.south);
        \node [circle, minimum size=14mm]       (gap) [right of = one]{\Large{$\cdots$}}
                edge    [fail, dashed] node[auto, swap] {}                      (one.north)
                edge    [recover, dashed] node[auto] {}                         (one.south);
        \node [state]   (minus one)     [right of = gap]                {$M-1$} 
                edge    [fail, dashed] node[auto, swap] {}                      (gap.north)
                edge    [recover, dashed] node[auto] {}                         (gap.south);
        \node [state]   (M)                     [right of = minus one] {$M$}
                edge    [fail] node[auto, swap] {$(N+1)\lambda$}                 (minus one.north)
                edge    [recover] node[auto] {$\mu$}                            (minus one.south);
        \node [state]   (plus one)      [right of = M]          {$M+1$} 
                edge    [fail] node[auto, swap] {$N\lambda$}    (M.north);     
\end{tikzpicture}
\caption{\label{MarkovChain}. Markov chain for the individual repair model.}
\end{figure}

The inclusion of repair yields a new Markov chain, shown in Figure~\ref{MarkovChain}.   The modeling of drive failures is the same as in Section~\ref{sec: no repair}. When a drive is repaired the system moves from state $i$ to state $i-1$.   If drives are repaired at a constant rate $\mu$, independent of the number of failed drives, the system moves from state $i$ to $i-1$ with \emph{effective repair rate} $\mu_i = \mu$. 

\subsection{Simple Case: \RAID{4/5}}
\label{sec: OneCheckDrive}
A simple case of this model is \RAID{4/5} which, in essence, have $N$ data drives and $M = 1$ check drive, and the failure of any two drives
is fatal. The states in this model are $0$, $1$ and $2$, and the transition
matrix is:
\begin{equation}
A=\left[\begin{matrix}
-(N+1)\lambda & \mu & 0\\
(N+1)\lambda  & -\lambda N-\mu& 0\\
0             &\lambda N & 0
\end{matrix}\right]
\end{equation}
The entry $R_{2,0}(z)$ of the resolvent of $A$ is of interest:
$$R_{2,0}(z) = {{\left(N^2+N\right)\,\lambda^2}\over{\left(N^2+N\right)\,
 \lambda^2\,z+\left(\left(2\,N+1\right)\,\lambda + \mu \right)\,z^2+z^3}}$$
By the Laurent series of $R_{2,0}(z)$ and~(\ref{eq: moment-fmla}) we find:
\begin{small}
  \begin{eqnarray*}
    m_1(\mathcal{T})&=& {{\left(2\,N+1\right)\,\lambda+\mu}\over{N(N+1)\,
        \lambda^2}},\\
    m_2(\mathcal{T})&=&{{\left(6\,N^2+6\,N+2\right)\,\lambda^2+\left(8\,N+4
 \right)\,\mu\,\lambda+2\,\mu^2}\over{\left(N^4+2\,N^3+N^2\right)\,\lambda
 ^4}}.
  \end{eqnarray*}
\end{small}
This result for $m_1(\mathcal{T}) =$ MTTDL is consistent with Plank in
\cite{Plank_NoMDL}. The variance may be computed using the formula
$\Var(\mathcal{T}) = m_2(\mathcal{T})-m_1(\mathcal{T})^2$ and is
useful when applying techniques such as the Tchebycheff inequality to
estimate the confidence interval for $\mathcal{T}$.  After some
simplifications
\begin{small}
  $$\Var(\mathcal{T})={{\left(2\,N^2+2\,N+1\right)\,\lambda^2+\left(4\,N+2
      \right)\,\mu\,\lambda+\mu^2}\over{\left(N^4+2\,N^3+N^2\right)\,\lambda^4
    }}.$$
\end{small}
Our technique generalizes easily to more check drives (e.g. 2, 3 and
4), but the expressions are too large to include in this paper. It is
worth noting that moment calculations do not require eigenvalues of
$A$. In contrast, calculating PDL$_t$, or equivalently,
$\exp(t\,A)$ is typically performed via diagonalization of $A$, and is
easily done using numerical techniques. Limited theoretical
results can be obtained, but they are beyond the scope of this paper.

\subsection{General Case}

For arbitrary $M$, the transition matrix is given by:

\begin{footnotesize}
\begin{equation*}
A=\left[\begin{array}{rrrrr r}
-\lambda_0 & \mu_1                 & 0            & \ldots         &0     & 0\\
\lambda_0  &-(\lambda_1+\mu_1) & \mu_2                 & \ldots   & 0     & 0\\
0          & \lambda_1            & -(\lambda_2+\mu_2)  & \ldots  & 0     & 0\\
0  & 0          &\lambda_2                                 &\ldots  & 0 & 0 \\
\vdots & \vdots & \vdots & \ddots & \vdots &\vdots \\
0  & 0          &0                                 &\ldots  & 0 & 0 \\
0  & 0          &0                                 &\ldots  &  \mu_M & 0 \\
0          & 0                    & 0                      & \ldots            &-(\lambda_{M}+\mu_{M}) &0        \\
0          & 0                    & 0                      & \ldots           & \lambda_{M}            & 0
\end{array}\right],
\end{equation*}
\end{footnotesize}
with $ \lambda_{j} = (T-j)\,\lambda$, and $\mu_{j}  = \mu$.  

As in Section~\ref{sec: no repair results}, we can numericaly compute PDL$_5$ for this model.  We use a repair rate of $\mu = \frac{1}{ 6 \textrm{ hours}}$ for our calculations.
\begin{figure}[hbtp]
	\centering
	\includegraphics[width = .45 \textwidth, viewport = 30 190 550 610, clip]{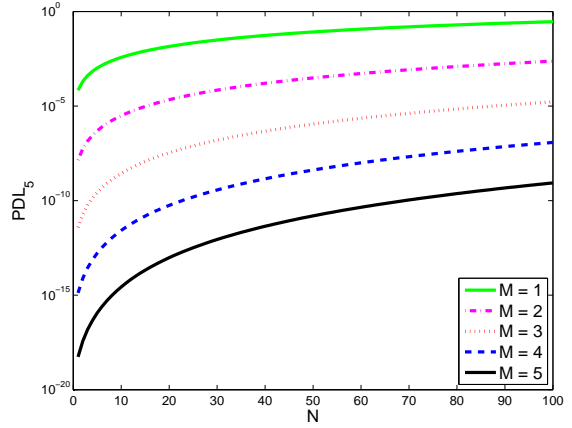} 
	\caption{\label{fig: simple results}PDL$_5$ for the individual repair model.}
\end{figure}

Figure~\ref{fig: simple results} shows PDL$_5$ as a function of $N$ for five values of $M$.  The $y$-axis (PDL$_5$) is on a logarithmic scale in this graph; each curve on the graph represents a particular value of $M$.  First, notice that these curves are spaced evenly apart for PDL$_5$ on a logarithmic scale.  This indicates that a RAID system under the individual repair model with $M+1$ check drives is \emph{exponentially} better than a RAID system with $M$ check drives and all other parameters the same.  That is, PDL$_5 \sim c_1^M$ where $c_1$ is a constant less than one.  Additionally, for a fixed $M$, the effect of increasing $N$ is relatively small.   
These observations have direct implications for the design of RAID systems.  Adding an additional check drive to a RAID will dramatically increase the reliability of that system, whereas an additional data drive will only decrease the reliability of the system slightly.  Therefore, the largest RAID systems are able to achieve the best reliability at the lowest cost per byte stored.


\section{Model 3: Simultaneous Repair \label{sec: sim repair}}
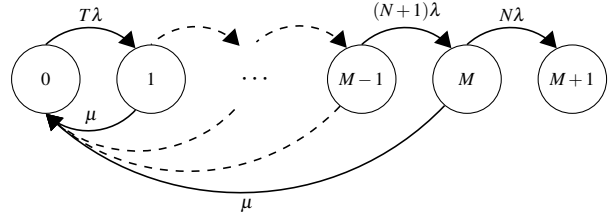
\begin{figure}[hbtp]
\centering
\begin{tikzpicture}[scale =.7, transform shape, node distance = 2 cm, bend angle = 45, > = triangle 60]
	\tikzstyle{state}=[circle, draw, minimum size=13mm]
	\tikzstyle{fail}=[<-, bend right, semithick]
	\tikzstyle{recover}=[->, bend left, semithick]
	\node [state] 	(zero) 						{$0$} ;
	\node [state] 	(one)  		[right of = zero]		{$1$}
		edge 	[fail] 		node[auto, swap] {$T\lambda$} 	(zero.north)
		edge 	[recover]	node[auto, swap] {$\mu$}				(zero.south);
	\node [circle, minimum size=14mm] 	(gap) [right of = one]{\Large{$\cdots$}}
		edge 	[fail, dashed] node[auto, swap] {} 			(one.north)
		edge 	[recover, dashed] node[auto] {} 			(zero.south);
	\node [state] 	(minus one) 	[right of = gap] 		{$M-1$} 
		edge 	[fail, dashed] node[auto, swap] {} 			(gap.north)
		edge 	[recover, dashed] node[auto] {} 			(zero.south);
	\node [state] 	(M) 			[right of = minus one] {$M$}
		edge 	[fail] node[auto, swap] {$(N+1)\lambda$}		 (minus one.north)
		edge 	[recover] node[auto] {$\mu$} 				(zero.south);
	\node [state] 	(plus one) 	[right of = M] 		{$M+1$} 
		edge 	[fail] node[auto, swap] {$N\lambda$} 	(M.north);	
\end{tikzpicture}
\caption{\label{fig: MarkovChain2} Markov chain for the simultaneous repair model.}
\end{figure}
As a revision to the individual repair model, it is unrealistic that hard drives are repaired one by one at some fixed rate.  Instead it is more likely that if one or more drives have failed, a repairman would be notified, and when he arrived to fix the drives, he would fix all failed drives at once.  This new model can be represented with the Markov chain shown in Figure~\ref{fig: MarkovChain2}. 
For this model, 

\begin{footnotesize}
\begin{equation*}
A=\left[\begin{array}{ccccc c}
-\lambda_0 & \mu_1                 & \mu_2            & \ldots       &\mu_M     & 0\\
\lambda_0  &-(\lambda_1+\mu_1)    & \ldots & 0                       & 0     & 0\\
0          & \lambda_1            & -(\lambda_2+\mu_2)  & \ldots &  0     & 0\\
0  & 0          &\lambda_2                                 &\ldots & 0 & 0 \\
\vdots & \vdots & \vdots & \ddots & \vdots &\vdots \\
0  & 0          &0                                 &\ldots &  0 & 0 \\
0  & 0          &0                                 &\ldots  &  0& 0 \\
0          & 0                    & 0                      & \ldots & -(\lambda_{M}+\mu_{M}) &0        \\
0          & 0                    & 0                      & \ldots &\lambda_{M}            & 0
\end{array}\right],
\end{equation*}
\end{footnotesize}
\noindent with $\lambda_m  = (T-m)\lambda$, and $\mu_m  = \mu$.
Notice for $M=1$ this is the same model as our first simple model, and has the same MTTDL.

\subsection{Results}
The effects on predicted reliability of the RAID as a result of this change to the model are negligible. For example, with $M = 5$, the PDL$_5$ for the simultaneous repair model is 0 - 3\% lower than for the individual repair model, and the effect grows linearly with $N$.  The same relationship holds for other values of $M$ with a smaller constant of proportionality for smaller $M$.  For large $M$ this effect might be significant but the reliability model is not sensitive to this modification.



\section{Model 4: Imperfect Repair} \label{sec: repair hazards}
Say you are a small company with a RAID system, and one drive or system fails.  You call in the appropriate employee, but this person may not be an expert in RAID.  He accidentally swaps out the wrong hard drive.  Now you effectively have a RAID system with two failed drives instead of one.

Here we will attempt to capture the effects of human error on the reliability of RAID systems.  We will build on the model discussed in Section~\ref{sec: sim repair}, using the same Markov chain (Figure~\ref{fig: MarkovChain2}) and transition matrix $A$, but will use different effective failure and repair rates.
We suggest that when hard drives fail there is a probability $p$ that in servicing those drives some other hard drive will be damaged and the already failed drives will not be repaired; there is a probability $1-p$ that the failed drives will successfully be repaired.  Therefore, the effective failure and repair rates are $\lambda_0 =  T\, \lambda$, $ \lambda_j =(T-j)\,\lambda + \mu\,p$ for $j > 0$, and $ \mu_j = \,\mu\,(1-p)$.

\subsection{Results}
\begin{figure}[htbp]
	\centering
	\includegraphics[width = .45 \textwidth, viewport = 30 190 550 610, clip]{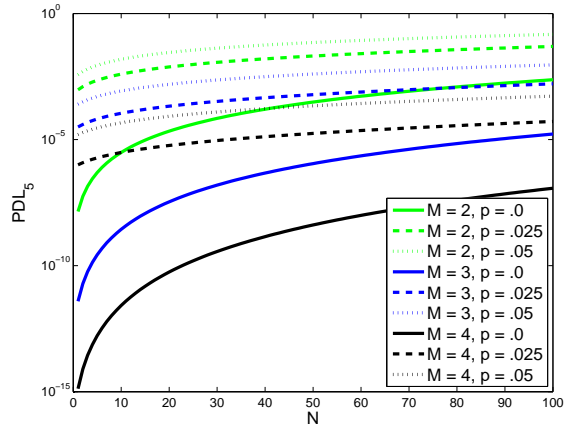}
	\caption{\label{fig: service results} PDL$_5$ for a model with imperfect repair for a variety of $M$ and $p$.  Notice that for $p = 0$, this is the simultaneous repair model.}
\end{figure}

Figure~\ref{fig: service results} shows the effects on PDL$_5$ of considering imperfect repair.  Notice that even for $p$ small, imperfect repair decreases the reliability of the system by several orders of magnitude.  Doubling $p$ decreases the reliability by at least one further order of magnitude.  For larger $M$ the effect is more pronounced with a decrease in reliability of as much as 10 orders of magnitude.  For someone who has designed their RAID system without considering the perils of service, they would need to add at least one if not two or more additional check drives to their RAID to maintain the same expected system reliability  in the face of service hazards.

\section{Model 5: Sector Errors}
Another major consideration in data reliability is the occurrence of irrecoverable read errors.  In particular, if a RAID system is rebuilding after $M$ drive failures, and encounters an irrecoverable read error on one of the remaining working disks, it will not be able to rebuild that byte for the RAID system.  As mentioned in Section~\ref{sec: intro}, irrecoverable read errors occur once in every $10^{14}$ bits, and thus are a common occurrence.
Here, we will restrict our attention to the particular case of sector errors, and assume that although sector errors are common it is unlikely that the same sector will fail on two or more disks simultaneously.  

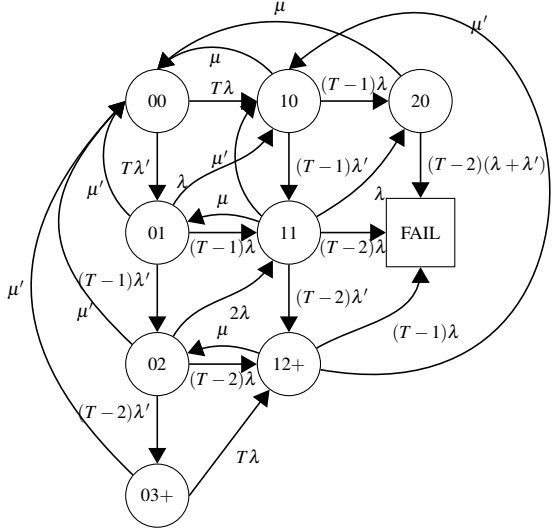
\begin{figure}[hbtp]
\centering
\begin{tikzpicture}[scale =.7, transform shape, node distance = 2.5 cm, bend angle = 45, > = triangle 60]
	\tikzstyle{state}=[circle, draw, minimum size=12mm]
	\tikzstyle{fstate}=[rectangle, draw, minimum size=13mm]
	\tikzstyle{fail}=[<-, semithick]
	\tikzstyle{fail2}=[<-, semithick, in = 60, out = -120]
	\tikzstyle{recover}=[->, bend right, semithick]
	\tikzstyle{recover2}=[->, semithick, in = 20, out = 160]
	\tikzstyle{scrub}=[->, bend left, semithick]
	\useasboundingbox (-3, -8) rectangle (8, 2);
	\node [state] 	(zerozero) 					{$00$} ;
	\node [state] 	(zeroone) 			[below of = zerozero] 	{$01$} 
		edge 	[fail]		node[auto] {$T \lambda'$}			(zerozero.south)
		edge 	[scrub]	node[near start, left] {$\mu'$}		(zerozero.west);
	\node [state] 	(zerotwo)  		[below of = zeroone]		{$02$}
		edge 	[fail]		node[near end, left] {$(T-1) \lambda'$}		(zeroone.south)
		edge 	[scrub, distance = 2.5cm] node[very near start, left] {$\mu'$}		(zerozero.west);
	\node [state] 	(onezero)  		[right of = zerozero]		{$10$}
		edge 	[fail] 		node[auto, swap] {$T\lambda$} 	(zerozero.east)
		edge 	[fail2] 	node[near end, left] {$\lambda$} 	 (zeroone)
		edge 	[recover]	node[auto] 	{$\mu$}			(zerozero.north);
	\node [state] 	(oneone) 			[right of = zeroone] 		{$11$}
		edge 	[fail] 		node[auto] {$(T-1)\lambda$} 	(zeroone.east)
		edge 	[fail]		node[auto, swap] {$(T-1) \lambda'$}	(onezero.south)
		edge 	[fail2] 	node[auto] {$2\lambda$} 	 		(zerotwo)
		edge 	[recover2]	node[auto, swap] 	{$\mu$}		(zeroone)
		edge 	[scrub]	node[auto] {$\mu'$}			(onezero.west);
	\node [state] 	(twozero)  		[right of = onezero]		{$20$}
		edge 	[fail] 		node[auto, swap] {$(T-1)\lambda$} 	(onezero.east)
		edge 	[<-, semithick, in = 30, out = -120] 	node[auto] {$\lambda$} 	 		(oneone)
		edge 	[recover]	node[auto, swap] {$\mu$}			(zerozero.north);
	\node [state] 	(zerothree)  		[below of = zerotwo]		{$03+$}
		edge 	[fail]		node[near end, left] {$(T-2) \lambda'$}(zerotwo.south)
		edge 	[scrub, distance = 3.5 cm]	node[auto] {$\mu'$}		(zerozero.west);
	\node [state] 	(onetwo)  			[right of = zerotwo]		{$12+$}
		edge 	[fail] 		node[auto] {$(T-2)\lambda$} 		(zerotwo.east)
		edge 	[fail]		node[auto] {$T \lambda$}			(zerothree.east)
		edge 	[fail]		node[auto, swap] {$(T-2) \lambda'$}	(oneone.south)
		edge 	[recover2]	node[auto, swap] 	{$\mu$}		(zerotwo)
		edge 	[->, distance = 7cm, semithick, out=-10, in = 40]	node[near end, right] {$\mu'$}			(onezero.north);
	\node [fstate] 	(fail) 			[right of = oneone] 		{FAIL} 
		edge 	[fail] 		node[auto] {$(T-2)\lambda$} 		(oneone.east)
		edge 	[fail,  out = -90, in = 30] node[auto] {$(T-1)\lambda$} (onetwo)
		edge 	[fail]		node[auto, swap] {$(T-2)(\lambda + \lambda')$}	(twozero.south);
\end{tikzpicture}
\caption{\label{fig: MarkovChainSectors}. Markov chain for a model of RAID reliability that considers the effects of sector errors for the $M = 2$ case.  In this model, state $ij$ indicates $i$ failed drives and $j$ working drives with sector errors.}
\end{figure}

We include the effects of sector errors on reliability by considering a two-dimensional Markov chain where the state $ij$ signifies that $i$ drives have failed, and $j$ drives have sector errors, and restrict our states to $i + j \leq M + 1$.    There is a special FAIL state which indicates that data has been lost and corresponds to either $M+1$ failed drives and no working drives with sectors errors or $M$ failed drives and one or more working drives with a sector error.  We denote the case where $i + j = M+1$ as $ij+$ meaning that $i$ drives have failed and $j$ or more drives have sector errors.  As in the simultaneous repair model of Section~\ref{sec: sim repair}, drives fail at rate $\lambda$, and are simultaneously repaired at rate $\mu$.  In addition, clean drives will develop sector failures at rate $\lambda'$, and we scrub the drives to remove the sector errors at rate $\mu'$.  This model was studied in \cite{Paris_Read_Errors}, and the approach here is similar but  generalized to an arbitrary number of check drives.   Figure~\ref{fig: MarkovChainSectors} depicts the Markov chain for the $M = 2$ cases, and Table~\ref{tab: sector rates} gives the transition rates between states in the general case.


\begin{table}[hbtp]
\begin{small}
\begin{tabular}{l|l|l|p{.75in}}
  \hline  \hline
  &Transition & Rate & Condition\\
  \hline   \hline
  \multicolumn{4}{c}{Conditions: $i+j\leq M$ and $i,\,j\ge 0$}\\
  \hline 
  1&$(i,j) \rightarrow (i+1,j)$ & $(T - i - j)\lambda$ &\\
  2&$(i,j) \rightarrow (0,j)$ & $\mu$ & $i \ge 1$\\
  3&$(i,j) \rightarrow (i,j+1)$ & $(T - i -j)\lambda'$ &\\
  4&$(i,j) \rightarrow (i,0)$ & $\mu'$ & $j \ge 1$\\
  5&$(i,j) \rightarrow (i+1,j-1)$ & $j \lambda$ & $j\ge 1$\\
  \hline
  \multicolumn{4}{c}{Conditions: $ i + j = M + 1$  and $i,\,j\ge 0$}\\
  \hline
  6&$(i,j) \rightarrow (i+1,j-1)$ & $(T-i) \lambda$ & $j\ge 1$\\
  \hline  \hline
\end{tabular}
\end{small}
\caption{\label{tab: sector rates}%
  Table of transition rates for the sector failure model. We write $(i,j)$ instead of $ij$.
  Type 1 is failure of a drive without sector error.
  Type 2 is replacement of all failed drives. Type 3 is sector failure. Type 4 is removal of all sector errors (scrubbing).  Type 5 is failure of a drive with a sector error.
  Type 6 is a failure of a drive in the special case where $i + j = M + 1$.  Since this case indicates that $j$ \emph{or more} drives have sector errors, the failure of any drive regardless of whether that drive has sector errors moves the system to the state $(i+1, j-1)$.  
%
}
\end{table}

Similar to our previous methodology, we can consider a vector $\mathbf{q}(t)$ that gives the probability of being in each state of our Markov chain as a function of time.  Since we have been considering states indexed by two numbers, it is necessary to relabel our states in some one dimensional manner, but the convention by which we do this is unimportant.  We can then use the transition rates given in Table~\ref{tab: sector rates} to write a transition matrix.  Even for small $M$, this matrix is too large to include here.  One can then calculate $\mathbf{q}(t)$ using (\ref{eq: gen state solution}).  For numerical calculations, we assume a sector failure rate of $\lambda'  = 1/ (2$ days) and a scrub rate of $\mu' = 1/(6$ hours), except where otherwise noted.  This scrub rate is unrealistically high; we will find that it does not matter.

\begin{figure}[htbp]
	\centering
	\includegraphics[width = .45 \textwidth, viewport = 30 190 550 600, clip]{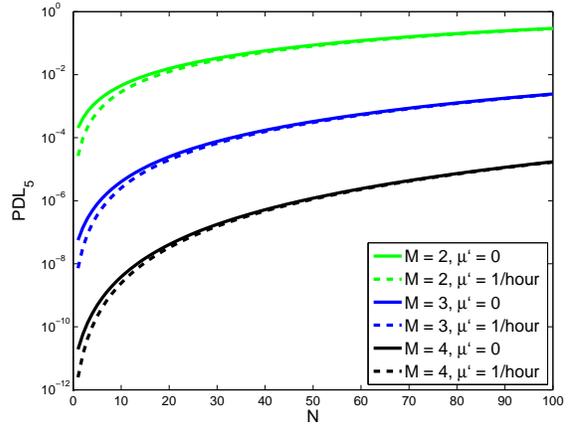}
	\caption{\label{fig: sector errors effect of scrubbing}PDL$_5$ for a model with sector errors.  This graph compares no scrubbing to scrubbing once an hour for $M = 2, 3$, and 4.}
\end{figure}

Figure~\ref{fig: sector errors effect of scrubbing} shows the PDL$_5$ for this model as a function of $N$ for various $M$ and $\mu'$.  Notice that for a fixed $\mu'$, this graph is qualitatively similar to Figure~\ref{fig: simple results}, the graph of PDL$_5$ for the individual repair model.  However this graph shows that the consideration of sector errors yields an estimate of PDL$_5$ that is several orders of magnitude higher than the model proposed in Section~\ref{sec: sim repair}.


\cite{Paris_Read_Errors} discusses extensively the benefits of shorter scrubbing intervals (increased $\mu'$), finding that more frequent scrubbing can substantially increase the MTTDL of the system and thus make it more reliable.  We find that scrubbing more frequently can indeed increase the reliability of the system, but the effect is much less dramatic than adding another check drive. Indeed, for the sector failure rate used, it was necessary to increase the scrub rate to once and hour to produce a discernible effect  on the graph in Figure~\ref{fig: sector errors effect of scrubbing}. 

\subsection{Sector Errors and Imperfect Repair \label{sec: sector imperfect repair}}


\begin{table}[hbtp]
\begin{small}
\begin{tabular}{l|l|p{1in}|l}
  \hline  \hline
  &Transition & Rate & Cond.\\
  \hline  \hline
  \multicolumn{4}{c}{Conditions: $i+j\leq M$ and $i,\,j\ge 0$}\\  
  \hline
  1a& $(0,j) \rightarrow (1,j)$ & $(T - j)\lambda$ & $j \leq M$ \\
  1b&$(i,j) \rightarrow (i+1,j)$ & $(T - i - j)\,\lambda + \frac{T-i-j}{T-i}\mu\,p$ & $i \ge 1$\\
  2&$(i,j) \rightarrow (0,j)$ & $\mu (1-p)$ & $i \ge 1$\\
  3&$(i,j) \rightarrow (i,j+1)$ & $(T - i -j)\lambda'$ &\\
  4&$(i,j) \rightarrow (i,0)$ & $\mu'$ & $j \ge 1$\\
  5a&$(0,j) \rightarrow (1,j-1)$ & $j \lambda$ & $j\ge 1$\\
  5b&$(i,j) \rightarrow (i+1,j-1)$ & $j \lambda + \frac{j}{T-i}\mu p$ & $j\ge 1$\\
  \hline
  \multicolumn{4}{c}{Conditions: $i+j=M+1$ and $i,\,j\ge 0$}\\  
  \hline
  6a&$(0,M+1) \rightarrow (1,M)$ & $T \lambda$ & \\
  6b&$(i,j) \rightarrow (i+1,j-1)$ & $(T-i) \lambda + \mu p$ & $ j \ge 1$\\
  \hline  \hline
\end{tabular}
\end{small}
\caption{\label{tab: sector rates imperfect repair}
  Table of transition rates for the sector failure model with imperfect repair. 
  As compared to Table~\ref{tab: sector rates}, there are a few differences.
  Types 1, 5 and 6 of of Table~\ref{tab: sector rates} split into two types, 1a-b, 5a-b and 6a-b respectively, in order to account for imperfect repair.
  The second terms of cases 1b and 5b represent the impact of imperfect repair which does not effect transitions from states where $i = 0$; the fractions $(T-i-j)/(T-i)$ and $j/(T-i)$ represent an erroneous
  repair of a functional drive without/with sector errors, respectively.
  }
\end{table}

\begin{figure}[hbtp]
	\centering
	\includegraphics[width = .45 \textwidth, viewport = 40 190 550 610, clip]{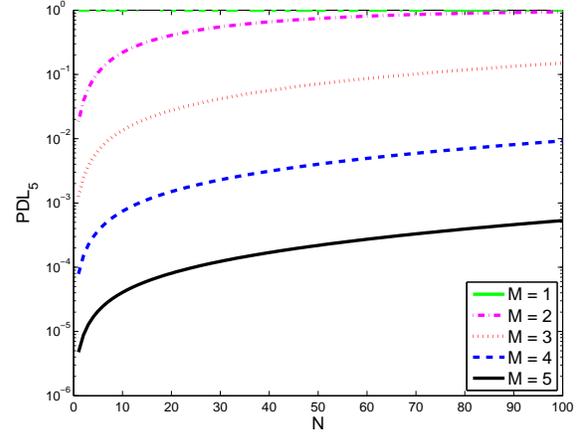}
	\caption{\label{fig: sector imperfect repair}PDL$_5$ for a model with with sector errors and imperfect repair with $p = 0.05$.}
\end{figure}

We can update our model of sector failures to also capture the effect of imperfect repair discussed in Section~\ref{sec: repair hazards} simply by altering the transition rates of the Markov model proposed in this section.   Table~\ref{tab: sector rates imperfect repair} shows the transition rates for the sector errors and imperfect repair model.  Figure~\ref{fig: sector imperfect repair} shows the PDL$_5$ for this model with $p = 0.05$.  Notice that the reliability estimates under this model are up to 6 orders of magnitude worse than for the model of sector errors without considering imperfect repair, and up to 14 orders of magnitude worse than the model without either sector failures or imperfect repair, for the values of $M$ shown.  

\section{Delay of Service}
We have just seen the hazards of service to the reliability of the system, so perhaps a way to mitigate these hazards is to plan to service the RAID system less frequently.  Section~\ref{sec: no repair} took this idea to the extreme, where we considered a system in which there is no service to hard drives.  Here we will consider a system where service happens, but might be delayed.  This could correspond to a situation where the RAID system is only serviced on weekends when their are fewer users and thus it might be more convenient for the system to be tied up in rebuild.  

So far, we have modeled the time to repair as an exponentially distributed random repair rate $\mu = 1/(6$ hours).  For a simple model of delayed repair, we will consider the effects on reliability of decreasing the rate of repair in the model proposed in Section~\ref{sec: sector imperfect repair} that considers sector errors and imperfect repair.

\begin{figure}[htbp]
	\centering
	\includegraphics[width = .45 \textwidth, viewport = 30 190 550 610, clip]{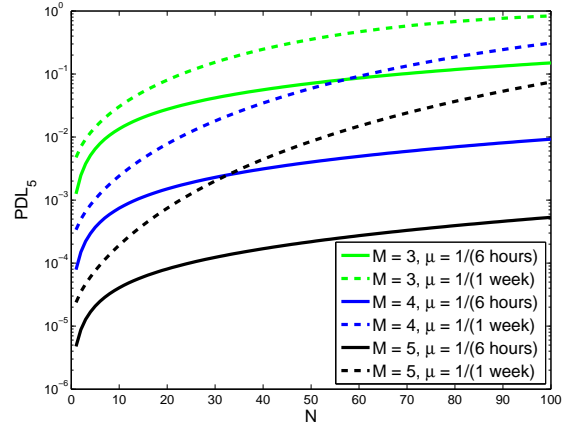}
	\caption{\label{fig: delayed service}PDL$_5$ for a model with with sector errors and imperfect repair.  This figure compares a mean time to repair of three hours to 1 week for $M = 3, 4, 5$.}
\end{figure}

Figure~\ref{fig: delayed service} compares the reliability of a RAID system that has a mean time to repair of 3 hours (the base model) to one with a mean time to repair of 1 week (delayed service) for $M = 3, 4, 5$.  The delay of service has a dramatic impact on the reliability of the system, particularly for large $N$.  If one intended to run a RAID system with delayed service, one would need to add one, two, or perhaps more additional check drives to compensate for the reduced reliability.

\section{Model 6: Delay Systems}
So far, we have modeled the time to repair and rebuild broken drives as a single, exponentially distributed random variable with rate $\mu$.   One can capture more nuance of the repair mechanism by modeling part of this process as a fixed time event requiring time $h$ to complete.  This might capture the significant fixed time required to repair and rebuild a RAID system, whereas the initiation of such service might happen at a rate $\mu$. We call this fixed time event a \emph{delay}.  For modern RAID systems,
due to the size and the number of drives, repair constitutes a significant portion of the 
system's usable life and thus delays may substantially impact our reliability model.


The mathematics of delay systems is substantially more complex than the models previously discussed.  To establish the theory, we begin with a toy example of a delay differential equation (DDE):
\[ 
	\frac{dy}{dt} = a\, y(t) + b\, H(t-1)\,y(t-1),\quad\text{$0\leq t< \infty$}, 
\]
where $H(t)=\int_{-\infty}^t\delta(t')\,dt'$ is the Heaviside step function.
Let $\hat{y}(s)=\int_0^\infty e^{-s\,t} y(t)\,dt$ be the Laplace transform of $y(t)$.
Using the usual calculus of Laplace transforms (i.e. integration by parts for derivatives
and change of variables for shifted arguments),
we obtain $s\,\hat{y}(s)-y(0) = a\, \hat{y}(s) + b\,e^{-s} \hat{y}(s)$.
Thus, $ \hat{y}(s) = y(0)/(s - a - b\,e^{-s})$.
Let $y(0)=1$ be the initial condition. Using the inverse Laplace transform (Mellin Inversion Formula) we get
\[ y(t) = \,\frac{1}{2\pi\,i}\lim_{\rho\to\infty}\int_{\sigma-i\,\rho}^{\sigma+i\rho}
\frac{e^{st}\,ds}{s-a-b\,e^{-s}} \] where $\sigma$ is chosen so that
all poles of the integrand lie to the left of the line $\Re
s=\sigma$, and is arbitrary, otherwise.  The poles of the integrand,
i.e. the roots of the equation $s-a=b\,e^{-s}$, are the
\emph{characteristic numbers} of the problem. The roots of the
equation $s\,e^{s} = z$ define the {\em Lambert $W$ function}, which
is a multi-valued function in the complex domain. Our equation is
$(s-a)e^s=b$ or $(s-a)e^{s-a}=b\,e^{-a}$. Hence, the characteristic
numbers are: $s= a+W(b\,e^{-a})$. Provably, there are infinitely
countably many values of $s$. If the path of integration can be
replaced with a large loop containing all characteristic numbers, and
if all characteristic numbers are simple roots
\footnote{All roots are simple unless $b\,e^{-(a-1)}=-1$.}
of the equation then, using the residue calculus, we find 
the \emph{formal solution}:
\[ y(t) = \sum_{j=1}^\infty \frac{e^{s_jt}}{1+be^{-s_j}} \]
where the summation is over all characteristic numbers $\{s_j\}$.
The full mathematical analysis of a system with
delay is subtle (such as convergence of the formal series above) and
will not be attempted here. It should be noted that numerical integration
of DDE's does not present any difficulty, and numerical results can
be readily obtained.

To obtain a model of RAID reliability, we must study systems of DDEs.  Here, we will consider only the following limited form:
\begin{equation}
  \label{eq: DDE}
\frac{d}{dt} \mathbf{q}(t) = B\,\mathbf{q}(t) + C\,H(t-h)\,\mathbf{q}(t-h)
\end{equation}
where $0\le t<\infty$, $B$ and $C$ are $n \times n$ matrixes and $\mathbf{q}(t)$ is a
vector-valued function with $n$ entries.
We take the Laplace transform and obtain:
\[ z\,\widehat{\mathbf{q}}(z) - \mathbf{q}(0) = B\,\widehat{\mathbf{q}}(z) + \exp(-z\,h)\,C\,\widehat{\mathbf{q}}(z) \]
where $\widehat{\mathbf{q}}(z)$ is the Laplace transform of $\mathbf{q}(t)$.
Hence, the Laplace transform of the solution is
\[ \widehat{\mathbf{q}}(z) = (z\,I-B-\exp(-z\,h)\,C)^{-1} \,\mathbf{q}(0)\]
where $R(z;B,C,h)=(z\,I-B-\exp(-z\,h)\,C)^{-1}$ is a complex,
matrix-valued, meromorphic function which replaces the ordinary resolvent for
a non-delay linear system. We note that for $h=0$ we have $R(z;B,C,h)
= R(z;A)$ where $A=B+C$ (the right-hand side is the ordinary
resolvent). The theory of moments of the solution discussed in Section~\ref{sec: Analytic MTTDL} carries through to
the delay case, and in particular involves only the Laurent series of
$R(z;B,C,h)$ at $z=0$. The computation of the probabilities, i.e. the
vector $\mathbf{q}(t)$, presents similar theoretical difficulties as
the toy example in the previous section. Numerical calculations are
straightforward. An analytical approach must deal with the poles
of $R(z;B,C,h)$, which are the roots of the transcendental characteristic equation
$\det\left(z\,I-B-\exp(-z\,h)\,C\right)=0$.

\subsection{Delay with One Check Drive}
Now we will modify the individual repair model for \RAID{4/5} discussed in 
Section~\ref{sec:  OneCheckDrive} to include delay.  We do so by assuming that disk repair takes no time (replacement of the
drive, reconstruction using data on functional drives, etc.), but
after that we simply wait $h$ units of time before adding the drive to
the RAID.  This is equivalent to assuming that repair takes time $h$ but that drives cannot fail during repair and reconstruction.
This model is not realistic and
serves only as an illustration of our approach.

\begin{figure}
  \centering
  \includegraphics[width =0.49\textwidth, viewport = 0 80 800 550, clip]{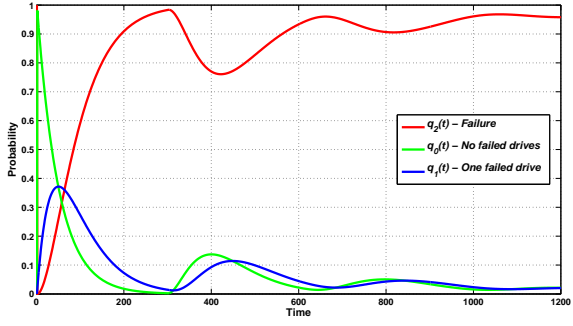}
  \caption{\label{fig: delay-oscillations} RAID with $M=1$, $T=2$, $\lambda=0.01$ and $\mu=0.01$ (both
    have units of inverse time); delayed repair of $h=300$ time
    units. MTTDL is the area above the graph of $q_2(t)$ as $0\le
    t<\infty$. According to our calculations, this area can be made
    arbitrarily large by increasing delay ($h$).}
\end{figure}

This model yields the system of differential equations:
\begin{small}
\begin{eqnarray}
  \label{eq: DDE_Example_1}
  \frac{d}{dt}q_0(t) &=& -(N+1)\,\lambda q_0(t) + \mu\,H(t-h)\,q_1(t-h),\\
  \label{eq: DDE_Example_2}
  \frac{d}{dt}q_1(t) &=&  (N+1)\,\lambda q_0(t) - (N\,\lambda\, + \mu)\,q_1(t),\\
  \label{eq: DDE_Example_3}
  \frac{d}{dt}q_2(t) &=&  N\,\lambda\, q_1(t).
\end{eqnarray}
\end{small}
The delay term represents disks for which repair started at time $t-h$
and are coming on-line at time $t$.
This system is of the form~(\ref{eq: DDE}), where
\begin{small}
\[ B=\left[\begin{array}{cc|c}
  -(N+1)\,\lambda & 0 & 0\\
  (N+1)\,\lambda  & -N\,\lambda -\mu& 0\\
  \hline
  0          & N\,\lambda & 0
  \end{array}\right],\quad
C=\left[\begin{array}{cc|c}
  0 & \mu & 0\\
  0  & 0 & 0\\
  \hline
  0          & 0 & 0
  \end{array}\right].
\]
\end{small}
The first moment of the time of failure is:
\begin{equation*}
  m_1(\mathcal{T})=
  {{\left(2\,N+1\right)\,\lambda+\mu}\over{\left(N^2+N\right)\,
      \lambda^2}}+{{\mu\,h}\over{N\,\lambda}}.
\end{equation*}
Clearly, when $h=0$, we reproduce the result of Section~\ref{sec:
  OneCheckDrive}. Unexpectedly, positive delay
results in larger $m_1(\mathcal{T})$. By delaying the drive replacement
we could in principle increase the mean time to failure indefinitely.  
This result is counterintuitive, and indicative of how unrealistic this model is.
In simulations, the state vector $\mathbf{q}$ exhibits oscillatory
behavior, as shown in Figure~\ref{fig: delay-oscillations}, which is
typical when studying delay phenomena.
The value of $\mathrm{PDL}_t=q_{2,0}(t)$ may also be obtained in principle via the inverse Laplace
transform, but due to the implicit dependency on the roots of the
characteristic equation, is not very tractable analytically.

\subsection{Delay with One Check Drive and Failure During Rebuild }
To improve the model of the previous section, we explicitly track
drive's progress while it is rebuilding, and allow it to fail during
rebuild.  The cost of this improvement is increased mathematical
complexity, as the model is a PDE. Under the notation of the previous
section, we assume that after the drive is replaced, it requires $h$
units of time to restore the data, but now the drive may fail during
recovery. The RAID may be in three functional states: no drives
failed, one drive failed and waiting to be replaced, and one drive
being reconstructed. In addition, there is the fourth, implicit state:
the failure state. The probability distribution consists of three
components:
\begin{enumerate}
\item $q_0(t)$ - the probability that a RAID system has no failed drives;
\item $q_1(t)$ - the probability that a RAID system has one failed drive waiting
  to be replaced;
\item $q_2(t,x)$ - the probability density of RAID systems with one drive
  being reconstructed; thus, the probability that a system
  is being reconstructed and the reconstruction has lasted between $x$ and
  $x+\Delta x$ units of time is $q_2(t,x)\,\Delta x$. Thus, $0\le x \leq h$.
\end{enumerate}
The system of differential equations (a generalization of Kolmogorov-Chapman
equations) for this system is:
\begin{small}
\begin{eqnarray}
  \label{eq: PDE_1}
  \frac{\partial q_0(t)}{\partial t} &=& -(N+1)\,\lambda\,q_0(t) + q_2(t,h),\\
  \label{eq: PDE_2}
  \frac{\partial q_1(t)}{\partial t} &=& (N+1)\,\lambda\,q_0(t) - (N\lambda + \mu)\, q_1(t),\\
  \label{eq: PDE_3}
  \frac{\partial q_2(t,x)}{\partial t} &=& -(N+1)\lambda q_2(t,x) -\frac{\partial q_2(t,x)}{\partial x},\\
  \label{eq: PDE_4}
  q_2(t,0) &=& \mu q_1(t).
\end{eqnarray}
\end{small}
This is a system of partial differential equations where the last equation is a boundary condition.
Most of the terms are familiar, except for the following terms:
\begin{enumerate}
\item $q_2(t,h)$ in equation~(\ref{eq: PDE_1}), which is a contribution from
  the drives for which reconstruction ended;
\item $-\frac{\partial q_2(t,x)}{\partial x}$ in equation~(\ref{eq: PDE_3})
  which is due to the RAID reconstruction progress;
\item Equation~(\ref{eq: PDE_4}) accounts for replacing failed drives, for
  which reconstruction begins immediately.
\end{enumerate}
Starting with all RAID systems with no failed drives and no 
drives being reconstructed, we arrive at the following initial conditions:
$q_0(0) = 1$, $q_1(0) = 0$, $q_2(0,x) = 0$ for $0\leq x \leq h$. It is possible to
obtain the MTTDL for this system by the methods previously introduced. The system
is solved by considering the Laplace transform in the time domain:
\begin{small}
\begin{eqnarray*}
  z\,\hat{q}_0(z) - q_0(0) &=& - (N+1)\,\lambda\,\hat{q}_0(z) + \hat{q}_2(z,h),\\
  z\,\hat{q}_1(z) - q_1(0) &=& (N+1)\,\lambda\,\hat{q}_0(z) - (N\lambda + \mu)\, \hat{q}_1(z),\\
  z\,\hat{q}_2(z,x) - q_2(0,x) &=& -(N+1)\lambda \hat{q}_2(z,x) -\frac{\partial\hat{q}_2(z,x)}{\partial x},\\
  \hat{q}_2(z,0) &=& \mu \hat{q}_1(z).  
\end{eqnarray*}
\end{small}
The strategy to solve the system is obvious: first we solve the third
equation as an ODE:
\begin{small}
\begin{eqnarray*}
  \hat{q}_2(z,x) &=& e^{-((N+1)\lambda+z)x}\,\hat{q}_2(z,0)\\
  &+& \int_0^x e^{-((N+1)\lambda + z)u}\, q_2(0,x-u)\,du.
\end{eqnarray*}
\end{small}
For given initial conditions, this reduces to $\hat{q}_2(z,x) =\mu\,
e^{-((N+1)\lambda+z)x}\hat{q}_1(z)$.  Next, we substitute this result into the first
equation and solve the resulting algebraic linear
system. If $\mathbf{q}(z)=(q_0(z),q_1(z))$ then the solution for given
initial conditions may be represented as $\mathbf{q}(z) =
(z\,I-A(z))^{-1}\mathbf{e}_0$, where $\mathbf{e}_0=(1,0)$ and
\begin{equation}
A(z) = \left[\begin{array}{cc}-(N+1)\lambda & \mu\,e^{-((N+1)\lambda+z)h}\\(N+1)\lambda &-(N\lambda + \mu)\end{array}\right].
\end{equation}
We have
\begin{small}
\[1-\mathrm{PDL}_t=P(\mathcal{T}> t)=q_0(t)+q_1(t)+\int_0^hq_2(t,x)\,dx.\]
\end{small}
The Laplace transform is $\hat{q}_0(z)+\hat{q}_1(z)+\int_0^h\hat{q}_2(z,x)\,dx$,
or explicitly:
\[\hat{q}_0(z)+\hat{q}_1(z)\\
+\frac{\mu}{(N+1)\,\lambda+z}\left(1-e^{-((N+1)\,\lambda+z)h}\right)\hat{q}_1(z).
\]
The $0$-th coefficient of the Taylor expansion with respect to $z$ yields:
\begin{small}
\begin{eqnarray*}
  \mathrm{MTTDL} &=&\frac{\left(2\,N+1\right)\,\lambda+\left(2-e^{-(N+1)\,\lambda\,h}\right)\,\mu}{
    \left(N^2+N\right)\,\lambda^2
    +\left(1-e^{-(N+1)\,\lambda\,h}\right)(N+1)\,\lambda\,\mu}\\
  &=&{{\left(2\,N+1\right)\,\lambda+\mu}\over{\left(N^2+N\right)\, \lambda^2}}
  -{{\left(\left(\mu\,N+\mu\right)\,\lambda+\mu^2\right)\,h}\over{N^2\,\lambda^2}}
  +\ldots.
\end{eqnarray*}
\end{small}
Again, for $h=0$ we reproduce the result of Section~\ref{sec:
  OneCheckDrive}. One can check that
$MTTDL$ is monotonically decreasing function of $h$ for all $h\geq 0$.
The limit as $h\to\infty$ is:
$${{\left(2\,N+1\right)\,\lambda+2\,\mu}\over{\left(N+1\right)\,
 \lambda\,\left(N\,\lambda+\mu\right)}}$$
As in the previous examples, obtaining explicit solutions for $\mathrm{PDL}_t$
presents a challenge, as it depends on the ability to solve a transcendental
equation akin to the Lambert equation.

We should note that the delay system~(\ref{eq: DDE_Example_1})-(\ref{eq: DDE_Example_3}) studied in the previous section
may be formally obtained from the system~(\ref{eq: PDE_1})-(\ref{eq: PDE_4}) by dropping the term $-(N+1)\lambda q_2(t)$ in equation~(\ref{eq: PDE_3}).

The technique introduced here is flexible enough to handle very
sophisticated RAID models, for instance, $M>1$ and the infamous
``bathtub curve'' (non-uniform failure rate of a drive, depending on
its age). However, presenting these results is beyond the scope of
this paper.

\section{Silent Data Corruption}
Occasionally, data is corrupted on the disk in a way that is not
detectable by hardware.  This incorrect data is then read and
delivered to the user as though it were correct.  An advantage of
Reed-Solomon codes is their ability to detect and correct such errors.
Reed-Solomon codes are an example of \emph{maximum distance separable}
codes, about which there are well known results in there area of
coding theory.  See \cite{Sudan_2001}, for example.

RAID systems group bytes on the drives into words, and words in the
same position on each data drive are combined to form the word in that
position on the check drive.  A RAID system with $M$ check drives and
all drives working can detect up to $M$ (and sometimes more) corrupt
words in any position.  Such RAID systems can correct $\frac{M}{2}$
corrupt words in a any position.

Given that silent data corruption occurs relatively infrequently
(see \cite{Panzer_2007} for a discussion of many types of errors), it
is sufficient to design RAID systems that are able to detect and
correct a single error at a time.  This requires $N+2$ drives to be
operational at any time.  If we think of PDL$_t$ as a function of $N$
and $M$ for one of the models previously discussed, then a system with
$N$ data drives and $M$ check drives has a probability of entering a
state where it cannot check for and correct silent data corruption of
PDL$_t(N+2, M-2)$ under that model.  Put differently, if you have
designed a RAID system that meets your reliability needs without
considering silent data corruption, then the addition of two check
drives allows you to check for and correct silent data corruption with
similar reliability.

\section{Conclusions}
\ifthenelse{\boolean{}}{
  \begin{itemize}
  \item discuss ability to schedule rebuild for a convenient moment - deferral of service.
  \item sensitivity of numerical results to parameter values.
  \end{itemize}
}{}

Given 100 drives, how do we design a RAID system to give the best reliability with the highest data rate (percent of drives used to store data)? We have studied the following possible designs:
\begin{small}
\begin{description}
\item[Double Mirroring] Each drive is mirrored twice.  This yields a data rate of 33/99.
\item[Two Independent \RAID{6}] Divide the 100 drives into two
  independent 50 drive \RAID{6} systems ($N=48$, $M=2$).
  If each RAID system has a PDL$_5$ of $q$,
  then the PDL$_t$ for the two systems together is PDL$_t = 1 - (1-q)^2$.
\item[Layered \RAID{6}] Divide 100 drives into 10 \RAID{6} groups.
  Let the 10 groups act together as a \RAID{6}.
  Let $m$ be the MTTDL for each ten drive group.
  Then we can model PDL$_t$ for the system by setting $\lambda = 1/m$.
  Note: for the sector failure model, we only model sector failures at the first level, not the second.
\item[Large RAID] A single large RAID system with $M=4$ or 11, and $N=100-M$.
\end{description}
\end{small}
Clearly, the single large RAID system is superior to other designs considered.
\begin{table}[hbtp]
  \begin{footnotesize}
    \begin{tabular}{p{.6in}|p{.15in}|p{.53in}|p{.53in}|p{.53in}}
      \hline\hline
      & & \multicolumn{3}{c}{PDL$_5$}\\
      \cline{3 - 5}
      &Data Rate&   Simultaneous  Repair & Imperfect Repair &   Sector Failures \& Imperfect Repair\\
      \hline\hline
      Double mirroring & $\frac{33}{99}$ &$4.64 \cdot 10^{-7}$&$ 1.17 \cdot 10^{-1}$&$ 4.61 \cdot 10^{-1}$\\
      \hline
      Two independent \RAID{6} & $\frac{96}{100}$ &  $5.46  \cdot 10^{-4}$ &  $1.13  \cdot 10^{-1}$&  $ 9.26 \cdot 10^{-1}$\\
      \hline
      Layered \RAID{6} & $\frac{64}{100}$ & $6.47 \cdot 10^{-23}$& $3.19 \cdot 10^{-4}$& $ 4.85 \cdot 10^{-3}$\\
      \hline
      Single Large RAID ($M=4$)& $\frac{96}{100}$ & $9.61 \cdot 10^{-8}$ & $1.98 \cdot 10^{-4}$ & $8.78 \cdot 10^{-3}$ \\
      \hline
      Single Large RAID ($M=11$)& $\frac{89}{100}$ & $3.62 \cdot 10^{-23}$  & $8.55 \cdot 10^{-13}$ & $1.52 \cdot 10^{-11}$\\
      \hline\hline
    \end{tabular}
  \end{footnotesize}
  \caption{Comparison of $\mathrm{PDL}_5$ for several RAID designs.}
\end{table}

\bibliography{Reliability}{}

\begin{thebibliography}{10}

\bibitem{Burkhard_93}
W.A. Burkhard and J~Menon.
\newblock {Disk array storage system reliability}.
\newblock In {\em Fault-Tolerant Computing, 1993. FTCS-23. Digest of Papers.,
  The Twenty-Third International Symposium on}, pages 432--441, June 1993.

\bibitem{Elerath_2007}
J.G. Elerath and M.~Pecht.
\newblock {Enhanced Reliability Modeling of RAID Storage Systems}.
\newblock In {\em Dependable Systems and Networks, 2007. DSN '07. 37th Annual
  IEEE/IFIP International Conference on}, pages 175--184, 2007.

\bibitem{Goodman_2006}
Roe Goodman.
\newblock {\em {Introduction to Stochastic Models}}.
\newblock Dover Publications, Inc., second edition, 2006.

\bibitem{Panzer_2007}
Bernd Panzer-Steindel.
\newblock {Data integrity}.
\newblock Technical report, CERN/IT, April 2007.

\bibitem{Paris_Read_Errors}
Jehan-Fran\c{c}ois Paris, Ahmed Amer, Darrell D.~E. Long, and Thomas J.~E.
  Schwarz.
\newblock {Evaluating the Impact of Irrecoverable Read Errors on Disk Array
  Reliability}.
\newblock In {\em Proceedings of the IEEE 15th Pacific Rim International
  Symposium on Dependable Computing (PRDC09)}, November 2009.

\bibitem{Paris_2008}
Jehan-Fran\c{c}ois Paris, Thomas J.~E. Schwarz, Darrel D.~E. Long, and Ahmed
  Amer.
\newblock {When MTTDLs Are Not Good Enough: Providing Better Estimates of Disk
  Array Reliability}.
\newblock In {\em Proceedings of the 7th International Information and
  Telecommunication Technologies Symposium (I2TS `08)}, December 2008.

\bibitem{Patterson_1988}
David~A Patterson, Garth Gibson, and Randy~H Katz.
\newblock {A case for redundant arrays of inexpensive disks (RAID)}.
\newblock {\em SIGMOD Rec.}, 17(3):109--116, 1988.

\bibitem{Plank_RS_Tutorial}
James~S. Plank.
\newblock {A Tutorial on Reed-Solomon Coding for Fault-Tolerance in RAID-like
  Systems}.
\newblock {\em Software -- Practice {\&} Experience}, 27(9):995--1012,
  September 1997.

\bibitem{Plank_Correction}
James~S. Plank and Y.~Ding.
\newblock {Note: Correction to the 1997 Tutorial on {R}eed-{S}olomon Coding}.
\newblock {\em Software -- Practice {\&} Experience}, 35(2):189--194, February
  2005.

\bibitem{Plank_NoMDL}
James~S. Plank, Kevin~M. Greenan, and Jay~J. Wylie.
\newblock {Mean time to meaningless: MTTDL, Markov models, and storage system
  reliability}.
\newblock In {\em HotStorage '10: 2nd Workshop on Hot Topics in Storage and
  File Systems}, Boston, June 2010.

\bibitem{Schroeder_2006}
Bianca Schroeder and Garth~A. Gibson.
\newblock {Disk failures in the real world: What does an MTTF of 1,000,000
  hours mean to you?}
\newblock In {\em Proceedings of the 5th USENIX Conference on File and Storage
  Technologies}, 2007.

\bibitem{Strang_1980}
Gilbert Strang.
\newblock {\em {Linear Algebra and Its Applications}}.
\newblock Academic Press, Inc., second edition, 1980.

\bibitem{Sudan_2001}
Madhu Sudan.
\newblock {Coding Theory: Tutorial and Survey}.
\newblock In {\em Proceedings of the 42nd IEEE symposium on Foundations of
  Computer Science}, pages 36--36, Washington, DC, USA, 2001. IEEE Computer
  Society.

\bibitem{Thompson_1981}
W.~A. Thompson~Jr.
\newblock {On the Foundations of Reliability}.
\newblock {\em Technometrics}, 23(1):1--13, January 1981.

\end{thebibliography}
\bibliographystyle{plain}

\end{document}